\documentstyle[aps]{revtex}            

\begin{document}

\draft

\title{Absence of the Mott transition in infinite dimensions}
\author{Kurt Fischer}
\address{
Department of Applied Physics,
University of Tokyo,
Tokyo 113-8656,
Japan}

\date{March 16, 1999}
\maketitle

\begin{abstract}
It is shown that the half filled Hubbard model does not display the
Mott metal-insulator transition, in the limit of infinite dimensions.
The proposed transition on the Bethe lattice is proved to be an artifact.
In the atomic limit, the model is always in the antiferromagnetic phase,
which can be described by the Weiss mean field model.

\end{abstract}

\pacs{PACS numbers: 71.10.Fd, 71.30.+h, 71.28.+d}


The competition between the kinetic energy and the Coulomb repulsion of 
electrons in a metal can cause a  metal-insulator transition~\cite{Buch-Mott}.
The basic model which should describe such a transition, is the 
one-band Hubbard model at half filling,
\begin{equation}\label{Hubbard-model}
H =\frac{t}{\sqrt{z}}  \! \sum_{\stackrel{(ij)}{\sigma= \uparrow,\downarrow}} 
    c^+_{i\sigma} c_{j\sigma}^{\phantom{+}} 
  + U \sum_i ( c^+_{i\uparrow} c^{\phantom{+}}_{i\uparrow} - \frac{1}{2}) 
      ( c^+_{i\downarrow} c^{\phantom{+}}_{i\downarrow} - \frac{1}{2})  
\end{equation}
where the hopping $\propto t$ is between (not necessarily only nearest) 
neighbors $(ij)$ on a lattice of coordination number $z$, and the constant 
chemical potential has been incorporated in the Hamiltonian. 
A theoretical understanding of the nature of the Mott transition is still 
lacking. 
The reason is that it will occur when the gain of kinetic energy due to 
hopping $\propto t$ equals the loss of Coulomb energy $\propto U$, where 
an expansion in either $t$ or $U$ is not permitted.
However, it was observed that the Hubbard model simplifies but remains
nontrivial in the limit of infinite coordination number $z$ or dimensions
$\propto z$~\cite{Metzner/Vollhardt-inf-dim}, so that, $1/z$ being small
already for three-dimensional lattices, it could serve as a starting point for 
an expansion in $1/z$.
As further evidence it was cited that the two extreme cases $t=0$ and $U=0$, 
are captured exactly in the limit of infinite 
$z$~\cite{Georges/Kotliar-impurity-inf-dim}.

In particular, it is taken for granted that the infinite dimensional Hubbard 
model shows a Mott transition at zero temperature and a finite critical value 
of $U$, that is, a transition between a paramagnetic metal and a paramagnetic
insulator, after frustrating the tendency to antiferromagnetic ordering. 
However, there is a controversy about the precise form of this transition%
~\cite{review-Georges,Noack/Gebhardt-inf-dim,Bulla-inf-dim,Logan/Nozieres-inf-dim,Kehrein-inf-dim,Schlipf-inf-dim}.
Specifically, there is the question whether the Fermi liquid quasi-particle 
spectral weight vanishes at the same value of $U$ as the Mott gap of the 
one-particle excitation spectrum opens.
In addition the results for the quasi-particle spectral weight differ 
significantly already away from the proposed transition.

It is the intention of this article, to show that the discrepancy in the 
results of the various approaches predicting the Mott transition in infinite 
dimensions, can be explained by the fact that they are ill-defined, and that 
this transition is absent.
The proof consists of four steps:

First it is recalled that there is no Mott gap, if the limit of infinite
dimensions is taken on a Bravais lattice~\cite{Georges/Krauth-inf-dim-1},
because its one-particle density of states has an infinite width.

Therefore, secondly, a proof for the existence of the Mott transition has been 
suggested for the Bethe lattice at infinite $z$, because its one-particle 
density of states would be semicircular and hence would have a finite 
width~\cite{Georges/Krauth-inf-dim-1,Rozenberg/Zhang/Kotliar-inf-dim}.
In addition, it has been argued that the same semicircular density of states
is realized for a lattice where the hopping matrix elements 
are distance {\em independent}, Gaussian distributed random 
variables~\cite{Noack/Gebhardt-inf-dim,Georges/Krauth-inf-dim-2},
to fully frustrate antiferromagnetic ordering.
However, it will be shown that the Bethe and the fully frustrated lattice,
have no well defined thermodynamical limit, and hence no well defined 
one-particle density of states.

Thirdly, it was argued that such a semicircular density of states, or more 
generally a  density of states of finite width,
can be assumed {\em ad hoc}, together with a momentum independent one-particle 
self-energy, to simulate the limit of infinite 
dimensions~\cite{review-Georges,Rozenberg/Zhang/Kotliar-inf-dim} without 
referring to a particular lattice, but retaining a finite bandwidth as in 
finite dimensions.
However, it will be proved here that no Hamiltonian with such properties 
exists, describing a limit of infinite $z$.

Finally, it will be shown that even the qualitative features of the obtained
one-particle excitation spectrum are incorrect, because 
the limit of infinite dimensions, in the paramagnetic phase, does {\em not} 
reproduce correctly the atomic limit $t\to 0$.
Even in the limit of infinite dimensions, the correct atomic limit of the
Hubbard model~(\ref{Hubbard-model}) is the Heisenberg antiferromagnet.
In particular this will invalidate the widely used second-order perturbation
theory~\cite{Zhang/Rozenberg/Kotliar-inf-dim}.

For the first point, consider an infinite dimensional Bravais lattice, 
whose one-particle density of states $\rho_0$ never 
vanishes~\cite{Mueller-Hartmann-inf-dim-1}.
Because the self-energy is momentum independent for infinite dimensions,
the local one-particle Green function
$G(i\omega_n)= \int \rho_0(\epsilon) G(i\omega_n , \epsilon) d\epsilon $
fulfills
\begin{equation}\label{G-local}
G(\omega - i0^+) = \int \frac{ \rho_0(\epsilon) }{ %
            \omega - \epsilon -\Sigma(\omega) - i0^+ } d\epsilon .
\end{equation}
If the one-particle excitation spectrum 
$ \rho(\omega) = \Im G(\omega - i 0^+)/\pi  $
of the local one-particle Green function $G$ would have a Mott gap, 
$G$ would be real valued, if $\omega$ lies in the gap.
For such values of $\omega$, except for possible isolated zeroes of $G$, 
the self-energy $\Sigma$ must be real as well so that 
\[
\rho(\omega) = \rho_0( \omega - \Re \Sigma(\omega) ) = 0 ,
\]
which is a contradiction.
Hence a Mott transition is absent on Bravais lattices in infinite dimensions.

For the second point, in order to have a Mott transition and to preserve
the finite bandwidth of finite dimensional lattices, usually the Bethe lattice
is used, or a fully frustrated lattice where the hopping matrix elements are 
distance {\em independent}, Gaussian distributed random 
variables~\cite{review-Georges,Noack/Gebhardt-inf-dim}.
These lattices have, however, no translational invariance:

To begin with the Bethe lattice, the reason is, that this lattice cannot be 
obtained as the thermodynamical limit of its finite counterpart, the Cayley 
tree of coordination number $z$ and 
linear dimension $N$, because its surface to volume ratio is for large $N$
and $z$, proportional to $1-1/z$~\cite{Eggarter-inf-dim}, hence
remaining finite in the limit of large $N$,
while for a usual Bravais lattice this vanishes as $z/N$.
For the Ising model, for example, this leads to a spurious phase transition
of the Bethe-Peierls type~\cite{Eggarter-inf-dim}.

Now, all derivations of the limit of infinite $z$ require at some
stage~\cite{review-Georges,Noack/Gebhardt-inf-dim} that there is a skeleton
expansion of the self-energy $\Sigma_{ij} (\omega)$ which becomes 
local $\propto \delta_{ij}$, and hence momentum independent. 
For this last step translational invariance of the lattice is required.
While for a  Bravais lattice of coordination number $z$ this holds 
independently of the boundary conditions, asymptotically in the sense,
\begin{eqnarray}\label{Sigma-trans-inv}
\Sigma_{ij} (\omega) - \delta_{ij} \Sigma_{ii}(\omega) &=& O(1/z) \nonumber \\
\Sigma_{ii} (\omega) - \Sigma_{jj}(\omega) &=& O(z/N) , i \neq j 
\end{eqnarray}
allowing for the thermodynamical limit, the right hand side of the last 
equation, is of the order one for a Cayley tree.
Hence the arguments~\cite{review-Georges,Buch-Economou} leading to 
a momentum independent self-energy and the semicircular density of states of a
Bethe lattice, are invalid, because it is
{\em not} allowed to take the thermodynamical limit from the 
outset~\cite{Buch-Ashcroft/Mermin}.
The same arguments can be applied to the fully frustrated lattice, because
there by construction, every lattice point is at the surface.

In order to show the third point, the self-energy is supposed to be momentum 
independent, {\em and} a density of states of finite width $D$ is assumed 
{\em ad hoc} to be realized by a Hamiltonian system, in infinite dimensions.
The half-filled, paramagnetic case is considered.
Then the Luttinger-Ward functional $\Omega$ describing the free energy of the
system per lattice site,
\begin{eqnarray}\label{LW-lattice}
\Omega &=& - T \sum_{n,\sigma} \int \rho_0(\epsilon) [
\ln G^{-1}( i\omega_n,\epsilon)  
+ ( i\omega_n - \epsilon) G ( i\omega_n , \epsilon)  \nonumber \\
&-& \sum_\nu \frac{1}{2\nu} \Sigma^{(\nu)} (i\omega_n) G(i\omega_n , \epsilon) 
   -1  ] d \epsilon
\end{eqnarray}
of the one-particle Green function $G$, is known~\cite{Luttinger/Ward} to 
have the saddle point property $\frac{\delta}{\delta G}\Omega = 0 $ if the
Dyson equation holds.
Here, $\Sigma^{(\nu)}$ denotes the $\nu$th order self-energy, in terms of
skeleton diagrams. 
At low temperatures, the first order correction to the ground state energy 
is therefore given by replacing $G$ by its 
$T=0$ value.
All but one of the $2\nu$ Matsubara sums in the $\nu$th order
self-energy $\Sigma^{(\nu)}$ can be replaced by its $T=0$ integral 
limits~\cite{LuttingerI}.
This can be performed in $2\nu$ ways, so that the last three terms in 
Eq.~(\ref{LW-lattice}) become 
$ ( i\omega_n - \epsilon - \Sigma(i\omega_n) )G(i\omega_n , \epsilon) -1 $, 
to cancel each other.
The remaining term, converted into a real integral, gives the first order 
correction to the ground state energy as $\hat{\Omega}(T) - \hat{\Omega}(0)$,
with
\begin{equation}\label{low-T}
\hat{\Omega} =   -\frac{2}{\pi} \int\! \int d\omega d\epsilon 
f(\omega) \rho_0(\epsilon) \arg G^{-1}_{T=0}( \omega-i0^+ , \epsilon) 
\end{equation}
where arg is defined such that $\arg(1) = 0$, and $f$ is the Fermi 
function~\cite{LuttingerI}.

It the sequel, the insulating phase is assumed to exist for sufficiently large
$U$ at zero temperature, which has been 
shown~\cite{Rozenberg/Zhang/Kotliar-inf-dim} to solve Eq.~(\ref{G-local}) 
for the finite width, semicircular $\rho_0$.
Then the one-particle excitation spectrum $\rho(\omega)$ has a gap $\Delta$,
around zero frequency, and the local Green function
\begin{equation}\label{G-local-spectral}
G(\omega-i 0^+) = \int \ \frac{ \rho(\epsilon) }{ \omega- i 0^+ - \epsilon} 
                   d\epsilon
\end{equation}
is real valued for $|\omega| < \Delta$ and has a Taylor expansion around
zero as $G(\omega) = -\alpha \omega/ \Delta^2  + \dots $
with positive $\alpha = \Delta^2 \int \rho(\epsilon) / \epsilon^2 d\epsilon $
of the order one.
Because $\rho_0$ has a finite width $D$, this and Eq.~(\ref{G-local}) imply 
for the self-energy for small $\omega$ and $|\Sigma(\omega) - \omega| >D$ the 
asymptotic behavior,
\begin{equation}
\Sigma ( \omega) = \Delta^2 / (\alpha \omega) + \dots \,  .
\end{equation}
Using the Dyson equation 
\begin{equation}
G^{-1}( \omega-i0^+ , \epsilon) 
= \omega-i0^+ - \epsilon - \Sigma(\omega-i0^+) ,
\end{equation}
then for small temperatures $T \ll \Delta^2/D, \Delta$ and frequencies
$\omega \stackrel{<}{\approx} T$
relevant for the first order low temperature 
correction to the free energy, $\arg G^{-1}$ is given by
\begin{equation}
\arg (- \frac{\Delta^2}{ \alpha \omega-i0^+} ) = - \pi \Theta(\omega) .
\end{equation}
Hence a {\em negative} entropy
\begin{equation}\label{low-T-explicit}
\Omega(T) - \Omega(0) =   2 T \ln 2 + \dots 
\end{equation}
results, which is a contradiction.
The physical reason is of course, that there is no density of states of finite 
width, for the infinite dimensional limit of a Bravais lattice:
In particular, the  Mott transition on the ``Bethe lattice'' is an artifact.
Any attempt to solve the infinite dimensional Hubbard model on the ``Bethe
lattice'' will produce spurious results, because the ``Dyson equation''
does not correspond to a real Hamiltonian system, in infinite dimensions.

As to the fourth point, it could be asked, if not at least the qualitative 
features of the one-particle excitation spectrum in the paramagnetic 
phase are reliable, because in the limit of infinite $z$,
the atomic limit $t\to 0$ is reproduced exactly?
In particular it is believed~\cite{review-Georges} that two
Hubbard bands around the energies $\pm U/2$ are emerging, as $U$ is increased.
The numerical procedures as well as second order perturbation 
theory~\cite{review-Georges} suggest this.
In addition, second order perturbation theory reproduces, at half filling,
the atomic limit of the local Green function exactly~\cite{review-Georges}, 
\begin{equation}\label{G-atomic-limit}
G(i\omega_n)^{-1} = i\omega_n - (U/2)^2/ i\omega_n 
= i\omega_n - \Sigma^{(2)} (i\omega_n)
\end{equation}
so it seems that it can be used as an interpolation scheme between the two
extremes $t=0$ and $U=0$, and as a starting point for an expansion around the 
atomic limit.

However, this argument contains a flaw: 
Eq.~(\ref{G-atomic-limit}) shows only that {\em because} the bare second order 
self-energy is exact, there is no well defined functional of the full
Green functions, in terms of skeleton self-energies:
The limit of infinite $z$, in the paramagnetic phase, does {\em not} reproduce 
correctly the atomic limit $t\to 0$.
The reason is that the ground state for $t=0$ is highly spin degenerate, so 
that there is no diagram technique~\cite{Buch-Abrikosov}, to justify
a momentum independent self-energy.
{\em Even} in infinite dimensions, this degeneracy must be lifted 
first, to arrive at the Heisenberg antiferromagnet~\cite{Metzner-inf-dim}.
In fact, if the Luttinger-Ward functional is expanded to second order in 
the skeleton self-energy~\cite{Mueller-Hartmann-inf-dim-2}, neither a Mott 
transition nor a precursor of Hubbard bands were observed for intermediate $U$,
in the paramagnetic phase.
This completes the proof that there is no Mott transition in infinite 
dimensions.

Finally, it is focused on the infinite dimensional
Heisenberg antiferromagnet as the effective Hamiltonian in the limit of small
$t/U$, of the half-filled Hubbard model in infinite dimensions, with exchange 
coupling $J \propto t^2/(Uz)$.
Its ground state on a hypercubic lattice is the Ne\'el 
state~\cite{Kennedy/Lieb/Shastry-inf-dim},
suggesting that the infinite dimensional limit of the Heisenberg 
antiferromagnet is the Ising model, but this has not been shown explicitly so 
far.
For a proof, the Dyson-Maleev 
representation~\cite{review-Manousakis-Heisenberg}
$S_i^+ = b_i^+(1-b_i^+b_i^{\phantom{+}})$, 
$S_i^- = b_i$ and $S_i^z = b_i^+b_i^{\phantom{+}} - 1/2 $ for the respective
{\em ferro}magnetic Heisenberg model (coupling $-J$) in terms of bosons is 
used, acting on its ground state as vacuum.
The unphysical states are projected out by an additional, large 
on-site interaction
$\propto \sum_i b_i^+ b_i^+  \! b^{\phantom{+}}_i \! b^{\phantom{+}}_i$.
Now the same arguments as for the fermionic 
system~\cite{Mueller-Hartmann-inf-dim-2} can be applied.
In contrast to the fermionic model, the one-particle (spin wave) terms scale 
as $1/z$, so that the spin-wave density of states collapses to a 
$\delta$ function, leaving only the ferromagnetic Ising model in the limit of 
infinite $z$, which transforms to the antiferromagnetic Ising model if the
sign of $J$ is reversed.
The Ising model itself reduces in infinite dimensions to the Weiss mean 
field theory.
Corrections to order $1/z$ for the Heisenberg model can be calculated by 
mapping it onto an effective spin-boson 
model~\cite{Kuramoto/Fukushima-inf-dim}.

To conclude, what physical picture emerges from the preceding discussion?
Clearly a Mott transition in the paramagnetic phase, as envisaged on the
basis of a limit of infinite dimensions, is absent.
The physical reasons are twofold: 
Firstly, although a generic {\em finite dimensional} Bravais lattice has of 
course a finite bandwidth, its proper infinite dimensional counterpart has not,
thus prohibiting a Mott transition.
This defect of the theory {\em cannot} be corrected by enforcing a
finite bandwidth in infinite dimensions, because the resulting ``Dyson''
equation does not describe a meaningful Hamiltonian.

Secondly, there is no well defined limit of infinite dimensions for the 
Hubbard model in the atomic limit in the paramagnetic phase.
The reason is, that the ground state for a vanishing kinetic energy is highly 
degenerate, so that for an expansion around the atomic limit, this
intrinsic spin degeneracy has to be lifted:
A paramagnetic insulating phase does not exist. 
In infinite dimensions, a simple Weiss mean field picture emerges.

Further work on the subject will concentrate on these long range ordering
processes, which in view of the arguments presented above, seem to dominate 
the Mott phenomenon in infinite dimensions.

The author is indebted to Prof. N.~Nagaosa, Dr. S. Murakami, and 
Dr. Chang-Ming Ho, for numerous discussions on the subject.
The financial support by the Japan Society for the Promotion of Science is 
kindly acknowledged.



\end{document}